# Enhanced skyrmion motion via strip domain wall

Xiangjun Xing,[1,*] Johan Åkerman,[2,3,†] and Yan Zhou[4,‡]

[1]*School of Physics & Optoelectronic Engineering, Guangdong University of Technology, Guangzhou 510006, China*

[2]*Department of Physics, University of Gothenburg, Fysikgränd 3, 412 96 Gothenburg, Sweden*

[3]*Material & Nano Physics, School of ICT, KTH Royal Institute of Technology, 164 40 Kista, Sweden*

[4]*School of Science & Engineering, The Chinese University of Hong Kong, Shenzhen, Guangdong 518172, China*

## ABSTRACT

When magnetic skyrmions move under spin orbit torque in magnetic nanowires, they experience a skyrmion Hall effect, which pushes them towards the nanowire edge where they risk being annihilated; this puts an upper limit on how fast they can be driven. However, the same magnetic multilayer harboring skyrmions can sustain a Néel-type strip domain wall along the nanowire length, potentially keeping the skyrmions separated from the edge. Here we study the interplay between current driven skyrmions and domain walls and find that they increase the annihilation current and allow the skyrmions to move faster. Based on the Thiele formalism, we confirm that the emergent longitudinal repulsive force and the modified energy landscape linked to the domain wall are responsible for the enhanced skyrmion motion. Furthermore, we identify that the longitudinal repulsive force emerges because of the broken axisymmetry in the local magnetization in front of the skyrmion. Our study uncovers key aspects in the interplay between two topological magnetic textures from different homotopy groups and may inspire new device concepts.

*xjxing@gdut.edu.cn
†johan.akerman@physics.gu.se
‡zhouyan@cuhk.edu.cn

## I. INTRODUCTION

Magnetic skyrmions are localized topological solitons that have a spin structure with integer topological charge [1–3]. Apart from the topological stability and accessible very small size, skyrmions exhibit emergent electrodynamics [4–11]. As such, magnetic skyrmions have attracted intense research activities in the past few years in the hope to bring novel spin-based data storage and information processing applications to market. Flowing magnetic skyrmions tend to deflect their trajectories from the current direction, experiencing what is called a skyrmion Hall effect, owing to a transverse Magnus force associated with the nonzero topological charge [12–14]. Accordingly, in confined magnetic nanostructures, e.g. nanowires, magnetic skyrmions usually move steadily along one of the two long edges, where the confining force and the Magnus force are balanced at moderate current densities [10,12,15]. However, once the current density surpasses a critical value, the skyrmions will touch the sample boundary and be annihilated [16–18].

Apart from Néel skyrmions [12,13,19–23], chiral Néel domain walls [24,25] can stably exist in magnetic multilayer nanowires with interfacial Dzyaloshinskii-Moriya interaction (iDMI) [26–28] and perpendicular magnetic anisotropy (PMA). Previous studies revealed that Slonczewski-type spin-transfer torque could drive the motion of Néel-type skyrmions and domain walls very efficiently [10,21,24,25,29,30]. A recent study [31] demonstrated that a Néel-type strip domain wall aligned along the nanowire length can be stabilized by the Slonczewski-type spin torque if the current used is not excessively large. Elongated strip domain walls in magnetic nanowires have proven robust magnonic waveguides that enhance spin wave transmission [32]. A method of controllably writing strip domain walls into magnetic nanowires has also been established [31]. Using the strip domain wall as a buffer layer, it appears possible to improve the dynamic behavior of magnetic skyrmions subject to Slonczewski-type spin torque.

In this work, we use micromagnetic simulations alongside theoretical modeling to the current-driven motion of magnetic skyrmions in a magnetic multilayer nanowire containing a strip domain wall (mediated skyrmions). For comparison, the motion of magnetic skyrmions in the same nanowire without including strip domain wall (bare skyrmions) is also considered. To ensure the general validity of our results, we examine a wide range of values of those material parameters that are sensitive to the multilayer interfacial condition [19,20,33]. Throughout the considered range of material parameters, we find that the skyrmion motion under Slonczewski spin torque is enhanced by the domain wall and the accompanying skyrmion Hall effect is suppressed. By virtue of the Thiele approach [12,20,34], we clarify the mechanism behind the observed behaviors. Our study opens a new paradigm for the interplay and manipulation of different topological magnetic textures.

## II. RESULTS

### A. Device structure, model, and simulations

The platform of this study is magnetic nanowires patterned from an ultrathin multilayer film, which has a HM1/FM/AO(HM2)-like structure to generate iDMI and PMA, where FM is a ferromagnetic layer, HM1 and HM2 represent heavy-metal layers with strong spin-orbit coupling, and AO stands for a metal oxide layer. Experimentally, the possible combinations of materials could be $Pt/Co/AlO_x$ [22,23,33], $Pt/CoFeB/MgO$ [13,21,24,35], $Ta/CoFeB/TaO_x$ [7,12,36], $Pt/Co/Ta$ [21] etc. Depending on the interfacial environment, layer thickness, and specific combination of materials, the interface-sensitive material parameters i.e. the iDMI and PMA can vary over a large range. Practically, magnetic skyrmions can be written into the multilayer nanowire through a local nanocontact spin valve or magnetic tunnel junction [37], and strip domain walls can be injected into the multilayer nanowire from the wire terminals using an established approach [31]. Overall, the architecture of an operational device is analogous to that in Ref. [31], but here we only consider straight magnetic nanowires and concentrate on the magnetic dynamics induced by Slonczewski spin torque. As usual, the Slonczewski spin torque is provided by a vertical spin current resulting from the spin Hall effect or a magnetic tunnel junction [7,10,21,23–25,29,30,31,35,38].

We perform micromagnetic simulations to find the solution to the formulated question by numerically integrating the extended Landau-Lifshitz-Gilbert equation with a spin-transfer torque [10,29,30],

$$\partial \mathbf{m}/\partial t = -\gamma (\mathbf{m} \times \mathbf{H}_{eff}) + \alpha (\mathbf{m} \times \partial \mathbf{m}/\partial t) + \mathbf{T}, \quad (1)$$

where $\mathbf{m}$ is the unit vector of the magnetization $\mathbf{M}$ normalized by its saturation value $M_s$ and $t$ is the time; $\mathbf{H}_{eff} = -(1/\mu_0)\delta E/\delta \mathbf{M}$ is the effective field in the FM layer with $\mu_0$ denoting the vacuum permeability and $E$ being the total energy incorporating the contributions of magnetostatic, PMA, exchange, iDMI, and Zeeman interaction terms; $\mathbf{T}$ represents the Slonczewski spin torque [39]; $\gamma$ is the gyromagnetic ratio and $\alpha$ is the Gilbert damping constant. For simplicity, we do not take into account the out-of-plane field-like torque. Also, the Zhang-Li torques were not included in our model since they are negligible even for the current-in-plane geometry [31].

The finite-difference code MuMax3 [40] was used to implement all the numerical calculations, in which only the FM layer is explicitly addressed. We do not directly incorporate the HM and AO layers in our simulations, but instead take account of the physical effects arising from them. In a real device, HM1 layer is responsible for the generation of spin currents in the FM layer via spin Hall effect and for the creation of iDMI together with AO or HM2 layer via forming asymmetric interfaces. The thickness of the FM layer, $d_{FM}$, is set to be 1 nm. The width of the nanowire is 100 nm in most cases and other values are also considered for special purpose. The nanowire length varies with the wire width but has a minimum of 1 μm. We examine the equilibrium magnetic configurations and their current-induced dynamics in a device with either or both of the skyrmion and strip domain wall over a broad range of $K_u$ and $D$. The presented results are based on the following material parameters unless otherwise specified: saturation magnetization $M_s$ = 580 kAm$^{-1}$, exchange stiffness $A$ = 15 pJm$^{-1}$, perpendicular magnetic anisotropy $K_u$ = 0.7 MJm$^{-3}$, iDMI strength $D$ = 3.0 mJm$^{-2}$, and Gilbert damping constant $\alpha$ = 0.3. These parameters are typical experimental values reported for the HM1/FM/AO(HM2) multilayer systems [7,12,13,21–24,33,35,36]. For computation, the FM layer is divided into an array of 1 × 1 × 1 nm$^3$ cubic cells, which are much smaller than the exchange length $l_{ex} = \sqrt{2A/\mu_0 M_s^2} \approx 8.4$ nm (the maximum length beyond which the short-range exchange interaction cannot keep all the magnetic moments parallel), and open boundary conditions are assumed.

In this study, we suppose that the Slonczewski spin torque stems from spin Hall effect [38], so that **T** = -$\gamma\tau_H$(**m** × **σ** × **m**), where $\tau_H = \hbar J\Phi_H/2e\mu_0 M_s d_{FM}$ and **σ** = $\hat{\jmath}$ × $\hat{z}$ being the spin current polarization direction with $\hbar$ denoting the reduced Planck constant, $J$ the electrical current density, $\Phi_H$ the spin Hall angle, $e$ the elementary charge, $\hat{\jmath}$ the unit vector in the electrical current direction, and $\hat{z}$ being the unit vector along the +$z$ axis. For dynamic simulations, the spin Hall angle $\Phi_H$ is set to be 0.13 [35,38] and the electric current in the nanowire is along –$x$. In the multilayer structure, we assume that the FM layer is on top of the HM1 layer. Then, the electrons' spin orientation **σ** = $\hat{\jmath}$ × $\hat{z}$ will orient along +$y$. For each pair of $K_u$ and $D$, the dynamic simulations are done for a series of current densities with an interval of 0.1 × 10$^{11}$ Am$^{-2}$.

## B. Domain wall dynamics

To control the skyrmion motion under spin orbit torque, the strip domain wall must be stable against the same torque. Therefore, first of all, we need identify the stability window of strip domain wall with respect to the driving electric current by multiple sets of simulations. Our calculations indicate that, for all the ($K_u$, $D$) combinations, the strip domain wall is not affected if the current density does not exceed $4.0 \times 10^{12}$ Am$^{-2}$, and otherwise it will collapse once a much larger current is applied. The two different situations are illustrated in Fig. 1.

Figures 1(a) and 1(g) display the initial static strip domain wall, which serves as the starting point of the dynamic simulations. In two separate simulations, the initial strip domain wall is supplied with lower and higher currents, respectively, and the subsequent temporal evolutions of the domain wall are recorded. The corresponding results are presented in Figs. 1(b-f) and 1(h-l). Clearly, at the lower current, the strip domain wall maintains its original profile; whereas at the higher current, the original narrow domain wall expands immediately after the current action, and meanwhile, its left end starts to divide into two branches. Rapidly, the strip domain wall develops a stripy substructure and the division extends deep into the interior of the nanowire.

As depicted in Figs. 1(a) and 1(g), for the applied current $I$, the electrons' spin orientation $\boldsymbol{\sigma}$ in the FM layer and the magnetization orientation $\mathbf{m}$ in the domain wall are parallel at the center of the domain wall, and thus the torque $\mathbf{T} \sim \mathbf{m} \times \boldsymbol{\sigma} \times \mathbf{m}$ vanishes therein regardless of the strength of the current density [31]. However, in the upper and lower magnetic domains, the magnetic moments are aligned along the $z$ axis and thereby the torque $\mathbf{T} \sim \mathbf{m} \times \boldsymbol{\sigma} \times \mathbf{m} = \hat{z} \times \hat{y} \times \hat{z} = \hat{y}$. When the current becomes considerably large, the torque will overcome the PMA and make the out-of-plane magnetic moments near the domain wall rotate to the $y$ axis, leading to the substructure inside the strip domain wall. Of course, once the current direction reverses, the strip domain wall will immediately collapse even for a small current density, because at this time the electron spins in the FM layer and the magnetic moments in the domain walls are in opposite directions, as demonstrated in Ref. [31].

The threshold value $J_{cd}$ = 4.0 × 10$^{12}$ Am$^{-2}$ is considerably large with regard to the skyrmion motion, since at $J_{cd}$ no skyrmions can stay in the nanowire for any ($K_u$, $D$). In fact, the maximal current density, which allows a skyrmion to stably exist in the nanowire, is slightly smaller than 1.0 × 10$^{12}$ Am$^{-2}$. This fact ensures that the strip domain wall can act as a tool for mediating the skyrmion dynamics.

It is worth noting that, according to an early paper [41], the Walker solution of a domain wall under an external field is unstable: a domain wall could move very slowly under a field, which is hard to detect in a micromagnetic simulation. If the same thing happens for a strip domain wall under a current, there will be a timescale beyond which the strip domain wall itself may disappear. However, as shown above and in our previous paper [31], the conventional Walker-type domain wall motion cannot happen in the present study. Therefore, the stability problem of a domain wall identified in Ref. [41] is safely avoided. In fact, the dynamics of a magnetic domain wall depends critically on the matching of domain-wall configuration and spin-orbit torques, as revealed in Ref. [29].

### C. Bare skyrmion dynamics

We check the current-induced skyrmion dynamics in the nanowire. Here, the starting point is a single static skyrmion. We carry out a series of simulations for each ($K_u$, $D$) to see the skyrmion dynamics under various current densities. In each simulation, the magnetization distributions are recorded as time sequences with a fixed temporal interval. These data allow us to identify how the skyrmion velocity depends on the current density, and, at which current density the skyrmion is annihilated due to skyrmion Hall effect. Two sets of representative results are shown in Fig. 2.

Figures 2(a) and 2(g) display the initial steady-state skyrmion that situates near the left end of the nanowire. At the time $t$ = 0 ns, an electric current is sent to the nanowire and then the skyrmion motion is initiated. Figs. 2(b-f) depicts the skyrmion dynamics for a lower subthreshold current density, at which the skyrmion moves through the nanowire and stops in front of the right edge. Whereas at a higher suprathreshold current density, the skyrmion's topological structure is destroyed when it contacts the sample boundary, as shown in Figs. 2(h-l). Eventually, this skyrmion is expelled from the nanowire. The dynamics of the bare skyrmion, demonstrated here, well agrees with what is known in previous research [10,15,17,18,30]. Apparently, in both cases, the skyrmion transverse displacement accompanies its longitudinal drift motion along the nanowire.

For the subthreshold current densities, the confining force from the sample boundary equilibrates with the Magnus force imposed by the current, and thereafter the skyrmion moves steadily along the nanowire. Nevertheless, for the suprathreshold current densities, the confining force is not strong enough to counteract the Magnus force, resulting in a net force that drives the skyrmion to move outward. According to the Thiele equation, the longitudinal skyrmion velocity $v_x$ and the Magnus force $F_g$ satisfy the relations $v_x \propto J$ and $F_g \propto J$, respectively [12,30]. In this context, the permitted maximal Magnus force determines the critical current density, which in turn defines the maximum skyrmion longitudinal velocity.

### D. Mediated skyrmion dynamics

To extend the upper theoretical limit of the skyrmion velocity, one has to suppress or avoid the skyrmion transverse motion. To this end, several classes of strategies have been proposed that use specially designed potential barriers [18,42], modified effective spin torque [17,43], or topologically compensated hybrid skyrmions, e.g., magnetic bilayer-skyrmions [16,22], antiferromagnetic skyrmions [44–47], and skyrmionium [48], to suppress the skyrmion Hall effect. These approaches can indeed give rise to increased skyrmion velocities; however, their realization requires rare materials, complex structures, and/or delicate operation. Especially, the adoption of antiferromagnetic skyrmions imposes a difficulty in the detection of information bits [22]. Therefore, other alternative ideas should be exploited for the development of fast spintronic devices. The skyrmion motion driven by the Slonczewski spin torque through the mediation of a strip domain wall manifests intriguing features, which are competing for use in spintronic technology and offer a basis for comprehending the dynamics of interacting magnetic textures.

Figures 3(a) and 3(g) show the coexisting skyrmion and strip domain wall in the steady state prepared for the dynamic study. In the following, two situations are considered: one corresponds to the subthreshold current densities [Figs. 3(b-f)], and the other to the suprathreshold current densities [Figs. 3(h-l)]. In either case, the skyrmion moves forward and simultaneously the strip domain wall maintains its major structure. Specifically, the skyrmion moves along the strip domain wall and just locally distorts the domain wall string. The whole process seems like a ball sliding along an elastic belt. The strength of the applied current distinguishes two kinds of dynamic behaviors. For a subthreshold current density, the skyrmion can safely pass through the nanowire with its size fixed, but for a suprathreshold current density, the skyrmion approaches the domain wall and contracts gradually, vanishing when its radius shrinks to zero.

Whether the skyrmion can move steadily with a stable size relies on if the confining force can cancel out the Magnus force experienced by the skyrmion. For the Magnus force on the mediated skyrmion, the relation $F_g \propto J$ [12,30] still holds, and accordingly, the higher the applied current, the larger the Magnus force. In this way, a larger current density will lead to a shorter distance and a stronger repulsive force between the skyrmion and domain wall. In principle, the threshold current density can be defined as the value at which the strip domain wall is maximally distorted by the skyrmion and meanwhile the skyrmion reaches its minimal stable size, and additionally the repulsive force is just able to offset the Magnus force. Once a bigger current is used, the Magnus force will continue to increase but the repulsive force will not, producing a nonzero net force that destroys the skyrmion. At smaller current densities, the repulsive force can always offset the Magnus force with the skyrmion stabilized in the transverse direction, enabling the steady drift motion of the skyrmion along the strip domain wall. In this respect, the strip domain wall serves to generate a confining force, playing the same role as the sample boundary. However, there are some fundamental differences between the strip domain wall and sample boundary, which will be discussed in the following sections.

### E. Skyrmion velocity versus current density

Now, we would like to describe the skyrmion motion quantitatively in terms of the skyrmion velocity versus current density (Figure 4). Without loss of generality, multiple sets of different ($K_u$, $D$) were considered. Fig. 4(a) shows the skyrmion velocity as a function of the current density for all the considered parameter combinations. For a direct comparison, the data are divided into two groups: one group is for the mediated skyrmion and the other for the bare skyrmion. Two striking characteristics are visible from this figure: the curves for the mediated skyrmion lie above those for the bare skyrmion and the upper curves extend to the higher-current density region. To make it clear, we plot the curves for each ($K_u$, $D$) in separate panels [Figs. 4(b-h)]. In each curve, the rightmost data point corresponds to the skyrmion motion at the current density just below the threshold value, above which the skyrmion cannot move steadily in the nanowire and will be annihilated. Then, the mentioned features of the curves reveal the following aspects: first, at an identical current density, the mediated skyrmion has a higher velocity than the bare skyrmion, and second, the mediated skyrmion can withstand stronger currents than the bare one. Consequently, the maximum velocity of the mediated skyrmion corresponding to the threshold current density is approximately twice that of the bare skyrmion at its own threshold current density irrespective of the ($K_u$, $D$), as shown in Figs. 4(b-h) and separately in Fig. 4(i).

The simulation results in Figures 1 to 4 substantiate that the strip domain wall can indeed act as a buffer layer to mediate the skyrmion dynamics, and, furthermore, the mediated skyrmion moves faster and permits using much stronger currents compared to the bare skyrmion. Nevertheless, these numerical results do not reflect what governs the observed behaviors. Next, we resort to the Thiele force equation to gain some insights.

## F. A Thiele model of the skyrmion motion

Assuming that the skyrmion has a rigid structure and projecting the extended Landau-Lifshitz-Gilbert equation onto the skyrmion translational mode, one obtains the generalized Thiele equation as follows [12,30],

$$\mathbf{G} \times \mathbf{v} - \alpha \overleftrightarrow{\mathcal{D}} \cdot \mathbf{v} + 4\pi B \overleftrightarrow{\mathbb{R}} \cdot \mathbf{J} + \mathbf{F}_p = 0, \quad (2)$$

which describes the balance of the Magnus force $\mathbf{F}_g$, dissipative force $\mathbf{F}_D$, driving force $\mathbf{F}_{ST}$, and confining force $\mathbf{F}_p$ acting on the skyrmion. In this work, we concentrate on the steady-state drift motion of a skyrmion along a nanowire, i.e., $\mathbf{v} = (v_x, v_y) = (v_x, 0)$. $\mathbf{G} = (0, 0, -4\pi Q)$ is the gyromagnetic coupling vector with the topological charge $Q = (1/4\pi) \int \mathbf{m} \cdot (\partial_x \mathbf{m} \times \partial_y \mathbf{m}) \mathrm{d}x \mathrm{d}y$, $\overleftrightarrow{\mathcal{D}} = \begin{pmatrix} \mathcal{D} & 0 \\ 0 & \mathcal{D} \end{pmatrix}$ is a dissipation tensor, $B$ quantifies the efficiency of the spin texture of a skyrmion absorbing the Slonczewski spin torque, and $\overleftrightarrow{\mathbb{R}} = \begin{pmatrix} \cos 0 & \sin 0 \\ -\sin 0 & \cos 0 \end{pmatrix}$ is an in-plane rotation matrix. $\mathbf{J} = (J, 0)$ is along the nanowire. Generally, $\mathbf{F}_p$ represents the force due to the confining potential associated with certain type of magnetic features such as boundaries, impurities, and magnetic objects [10,12,15,21,23,31]; here we intentionally assume that it incorporates two in-plane components, i.e., $\mathbf{F}_p = (F_x, F_y)$. Then, substituting these quantities into the vector equation (2), one finds,

$$\begin{cases} -\alpha \mathcal{D} v_x + 4\pi BJ + F_x = 0, \\ -4\pi Q v_x + F_y = 0 \end{cases} \quad (3)$$

After some simple algebra, one gets,

$$\begin{cases} F_y = \frac{4\pi Q}{\alpha \mathcal{D}}(4\pi BJ + F_x), \\ v_x = \frac{1}{\alpha \mathcal{D}}(4\pi BJ + F_x) \end{cases} \quad (4)$$

For the steadily moving bare skyrmion, the confining force due to the sample boundary is simply along the $y$ axis, pointing from the sample boundary to skyrmion, i.e., $\mathbf{F}_p = (0, F_y) = [0, F_p^{\perp}(\text{bSK})]$. Thus, for the bare skyrmion, one has,

$$\begin{cases} F_p^{\perp}(\text{bSK}) = \frac{4\pi Q}{\alpha \mathcal{D}} 4\pi BJ, \\ v_x(\text{bSK}) = \frac{1}{\alpha \mathcal{D}} 4\pi BJ \end{cases} \quad (5)$$

For the steadily moving mediated skyrmion, the confining force no longer simply points to the $y$ axis as for the bare skyrmion, and instead it has both $x$ and $y$ components, i.e., $\mathbf{F}_p = (F_x, F_y) = [F_p^{\parallel}(\text{mSK}), F_p^{\perp}(\text{mSK})]$. Then, for the mediated skyrmion, one sees,

$$\begin{cases} F_p^{\perp}(\text{mSK}) = \frac{4\pi Q}{\alpha \mathcal{D}}[4\pi BJ + F_p^{\parallel}(\text{mSK})], \\ v_x(\text{mSK}) = \frac{1}{\alpha \mathcal{D}}[4\pi BJ + F_p^{\parallel}(\text{mSK})] \end{cases} \quad (6)$$

It is easily noticed that, for the same current density $J$, $v_x(\text{mSK}) = v_x(\text{bSK}) + \frac{1}{\alpha \mathcal{D}} F_p^{\parallel}(\text{mSK})$. This result explains one of the main numerical findings, namely, the mediated skyrmion has bigger velocities than the bare one (Fig. 4).

For the steadily moving mediated skyrmion, the longitudinal repulsive force originates from the asymmetric distortion in the strip domain wall [refer to Fig. 5(a)]. Such asymmetric distortion destroys the axisymmetric local magnetization distribution with respect to $y$, enabling the emergence of an $x$ component in the repulsive force. However, for the steadily moving bare skyrmion, the local magnetization distribution from the skyrmion to the sample boundary always keeps axisymmetric relative to $y$, when the skyrmion approaches to the boundary, not allowing the existence of a net $x$-directed component in the repulsive force. As a result, the repulsive force arising from the boundary always orients along $y$ [Fig. 5(b)]. The detailed mechanism for the formation of the asymmetric domain wall distortion and the creation of the longitudinal repulsive force is clarified, as shown in the Supplemental Fig. S1 [49].

Generally, the forces exerted on a skyrmion by the physical boundary can be introduced in a first approximation as $F_p^{\perp}(\text{bSK}) = F_y = -k(y - y_0)$ [10,50], where $k > 0$ and $y_0$ is the skyrmion equilibrium position along the $y$ axis, by assuming a harmonic potential. For the forces imposed on a skyrmion by the strip domain wall, we assume that the above approximation still holds. Then, it is easy to obtain $F_p^{\perp}(\text{mSK}) = F_y = -\beta(\eta - \eta_0)$, where $\beta > 0$ and $\eta = y_{SK} - y_{DW}$ represents the interval along the $y$ axis between the skyrmion ($y_{SK}$ denotes the $y$-coordinate of the skyrmion center) and the strip domain wall ($y_{DW}$ signifies the $y$-coordinate of the bottom of the bent domain wall) after a certain current is applied, and $\eta_0 = \eta|_{J=0}$ represents the initial equilibrium interval without the current application.

Equation (6) suggests that $F_p^{\perp}(\text{mSK}) - \frac{4\pi Q}{\alpha \mathcal{D}}\left[F_p^{\parallel}(\text{mSK})\right] = \frac{(4\pi)^2 QB}{\alpha \mathcal{D}}J$, from which it is reasonable to suppose that $F_p^{\parallel}(\text{mSK})$ follows a similar relation, i.e.,

$$F_p^{\parallel}(\text{mSK}) = F_x = -\zeta(\eta - \eta_0), \quad (7)$$

where $\zeta > 0$. Now, letting $\Delta v_x = v_x(\text{mSK}) - v_x(\text{bSK})$, one has $\Delta v_x = \frac{1}{\alpha \mathcal{D}}F_p^{\parallel}(\text{mSK})$. Considering that $\mathcal{D} = \frac{\pi^2 d_{SK}}{8\lambda_{DW}}$ (where $d_{SK}$ and $\lambda_{DW}$ are the skyrmion diameter and domain wall thickness, respectively) [12], one finally obtains the following formula,

$$\Delta v_x = -\frac{8\zeta\lambda_{DW}}{\alpha\pi^2 d_{SK}}(\eta - \eta_0), \quad (8)$$

where $\lambda_{DW} = \sqrt{A/K_{eff}}$ with $K_{eff} = K_u - \frac{1}{2}\mu_0 M_s^2$ can be calculated directly from the material parameters, and $d_{SK}$, $\eta$, and $\eta_0$ can be derived from the simulation results. Leaving $\zeta$ as the free parameter, we fit the simulated velocity difference between the mediated and bare skyrmions (shown in Fig. 4) using Eq. (8). The fitting results are presented in Fig. 6, which contains five sets of data corresponding to various combinations of $K_u$ and $D$. Overall, the agreement between theory and simulations is good, considering that there exists only one free parameter; this fact indicates that the assumption of the form of the longitudinal force [i.e., Eq. (7)] is a very good approximation.

From Eqs. (5) and (6), for the same current density $J$, $F_p^\perp(\text{mSK}) = F_p^\perp(\text{bSK}) + \frac{4\pi Q}{\alpha D}F_p^\parallel(\text{mSK})$, suggesting that the repulsive force imposed by the strip domain wall is stronger than that exerted by the sample boundary. In fact, the confining potential on the mediated skyrmion can be much larger than the one on the bare skyrmion, which makes the mediated skyrmion be able to withstand a much stronger Magnus force. The mechanisms are clarified from comparing the ways of annihilation of the mediated and bare skyrmions. For the annihilation of the mediated skyrmion, there exist three optional routes. 1) The skyrmion could at first push a portion of the strip domain wall out of the boundary and then leave the sample from the boundary. In this case, because the strip domain wall is an extended entity, when even a piece of it approaches to the boundary, a large number of magnetic moments will join the strong local interaction magnetostatically causing a very strong repulsive force between the domain wall center and sample boundary [Fig. 5(c)]. Consequently, driving the domain wall to touch the boundary must overcome a huge energy barrier linked to the strong repulsive force. 2) The skyrmion may penetrate the strip domain wall and merge into the magnetic domain. However, the extending character of the strip domain wall together with the self-locking feature of the spin configuration between the skyrmion border and domain wall center lead to a still high energy barrier [Fig. 5(c)]. 3) The skyrmion may shrink gradually by contracting its border and vanish finally by absorbing a magnetic singularity [10,51]. Owing to the relatively small size of a skyrmion, the energy barrier associated with its annihilation is the lowest among the three situations. Therefore, a mediated skyrmion is always seen to annihilate through the route 3.

Nevertheless, only two possible pathways exist for the annihilation of the bare skyrmion. 4) As in the route 3 for the mediated skyrmion, the bare skyrmion could also be annihilated by shrinking its size and then absorbing a magnetic singularity. Here, however, the energy barrier is relatively high owing to the topological protection of the skyrmion and the requirement to inject a topological singularity [10,51]. 5) Alternatively, the bare skyrmion could be annihilated by touching the boundary. In this case, unlocking of the spin configuration between the skyrmion border and sample boundary can be simply launched by the reversal of those magnetic moments situating on the boundary [Fig. 5(d)], such that the entire skyrmion is easily erasable; the associated energy barrier is small. Thus, a bare skyrmion tends to be annihilated through the route 5.

Obviously, the three annihilation routes for the mediated skyrmion require overcoming larger energy barriers compared with the two annihilation routes for the bare skyrmion, determining that the mediated skyrmions have higher annihilation current densities than the bare ones and naturally can experience stronger Magnus forces.

### III. DISCUSSION

To check the stability of the mediated skyrmion motion, we numerically study the process in longer magnetic multilayer nanowires. The computational results indicate that the mediated skyrmion can propagate steadily over a considerably large distance with the shape and size fixed, as shown in Fig.

S2 [49]. We also consider the current-induced dynamics of an array of mediated skyrmions (Fig. S3 [49]) and found that the entire array moves concertedly when the interval is adequately large or the applied current is exceedingly low. Compared with the bare-skyrmion array, the smallest interval between two adjacent mediated skyrmions, which permits orderly motion, is larger, because, for the mediated-skyrmion array, a skyrmion is readily affected by its neighbors through bending of the domain wall. The results presented in this study do not rely on special material parameters and are universally valid for the HM1/FM/AO(HM2)-like multilayer system. As an example, the mediated skyrmion motion for a different ($K_u$, $D$) is displayed in Fig. S4 [49], where the entire process is essentially the same as in Fig. 3. A skyrmion, once driven to move, will adjust itself to a moderate stable size before reaching the steady-state motion; this is especially clear for a big skyrmion as shown in Fig. S4 [49]. These manifested characteristics of mediated skyrmions form the basic prerequisite for any realistic implementation of a device using them.

The extended Thiele equation including $F_p^{\parallel}$ provides insight into the numerical results, both qualitatively and quantitatively; it clearly demonstrates that the longitudinal repulsive force functions as an active driving force for the mediated skyrmion. According to Refs. [50,52], an inertia term is expected to enter the Thiele equation because of the edge confining potential. Since the strip domain wall also imposes a confining potential on the skyrmion as the physical borders, the effect of the mass term must have been involved in our simulation results. Despite the absence of the inertia term in our theoretical model, the agreement between the theory and simulations seems good overall, as demonstrated in Fig. 6. Therefore, in our opinion, the influence of the mass term is negligible in this scenario, and the model without considering the inertia has captured the key physics in the dynamics of the mediated skyrmion.

The proposed use of mediated skyrmions can suppress skyrmion Hall effect, namely, increasing the skyrmion mobility and expanding the effective working range of the current, and eliminate the random scattering of edge roughness on skyrmion motion. Nevertheless, it cannot avoid the skyrmion Hall effect, and thus there still exists a threshold current density $\sim 1.0 \times 10^{12}$ Am$^{-2}$, above which the mediated skyrmion will be annihilated. Analogously, a threshold current density also exists in most of the previously suggested schemes [16,18,42,48]; when the employed current density becomes exceedingly large, the skyrmions will be destructed by the uncompensated Magnus force. Actually, a recent literature [44] argued that the skyrmion Hall effect will still occur in the case of spin-polarized currents even for the skyrmions in antiferromagnets. Comparatively, our proposed scheme has remarkable advantages: First, it simply requires writing a strip domain wall into the original skyrmion device without incorporating the fabrication of complex hard structures and is thus naturally reconfigurable. Second, apart from skyrmionic devices, the hardware is also applicable to domain wall racetrack devices [53] and magnonic waveguides [32], without significant variation in the key parts, implying good reprogrammability.

Although the scattering on skyrmion motion by edge pinning sites can be prevented by using strip domain wall, our preliminary numerical calculations suggest that the impact of pinning centers [7,12,21,23] in the interior region of the nanowire is unavoidable, since the randomly distributed pointlike impurities pin the strip domain wall locally and modify its profile (Supplemental Fig. S5 and Movies S1–S3 [49]).

While the nanowire width is not crucial for the steady-state motion of a mediated skyrmion, the distance between the strip domain wall and sample boundary has a decisive role. An increased spacing will result in enhanced distortion of the strip domain wall and change the relative strength of $F_p^{\parallel}$ and $F_p^{\perp}$. For instance, in wider samples with a strip domain wall situating along the central axis, the steadily moving mediated skyrmion acquires higher velocities (as shown in Fig. 4c), because the heavier local bending of the strip domain wall, permitted by the bigger spacing between the domain wall and sample edge, results in a larger longitudinal component of the repulsive force. Fortunately, one can displace the strip domain wall using, for example, a magnetic field to reach an appropriate position.

Different from that of bare skyrmions, the motion of mediated skyrmions under spin orbit torque is unidirectional. A reversed direction of the applied current will at first cause the strip domain wall to deform randomly, and then the chaotic domain wall dynamics destructs the skyrmion leading to erroneous operation.

## IV. CONCLUSION

In conclusion, we point out theoretically the possibility to control current-induced skyrmion dynamics utilizing a strip domain wall. Through micromagnetic simulations, we study the dynamics of strip domain wall, bare skyrmion, and coexisting skyrmion and strip domain wall under spin orbit torque over a wide range of interface-sensitive material parameters. The computational results attest our theoretical conjecture and suggest that the skyrmion mediated by strip domain wall becomes faster and more stable, which is explained by the generalized Thiele equation with a two-component confining force. A symmetry analysis reveals that the longitudinal component of the confining force originates from local asymmetric distortion of the strip domain wall. The design of skyrmionic devices might benefit from these discoveries. More importantly, the study implies that, overall, the Thiele equation is robust in describing the dynamics of magnetic solitons [10], and specifically, the skyrmion velocity and Magnus force can be harnessed by a longitudinal force regardless of its origin.

## ACKNOWLEDGEMENTS

X.J.X. acknowledges support from the National Natural Science Foundation of China (Grants No. 11774069). Y.Z. acknowledges support by the President's Fund of CUHKSZ, the Longgang Key Laboratory of Applied Spintronics, the National Natural Science Foundation of China (Grants No. 11974298 and No. 61961136006), the Shenzhen Fundamental Research Fund (Grant No. JCYJ20170410171958839), and the Shenzhen Peacock Group Program (Grant No. KQTD20180413181702403).

**Author Contributions:** X.J.X. initiated and designed the study. Y.Z. coordinated the project. All authors contributed to the analysis of the results and wrote the manuscript.

## APPENDIX A: MICROMAGNETIC SIMULATIONS

The public-domain micromagnetic codes MuMax3 [40] is used to implement the micromagnetic simulations, in which the Landau-Lifshitz-Gilbert equation is numerically integrated, by means of the explicit Runge-Kutta method with an adaptive time step, to find the equilibrium magnetic configurations and trace the dynamics of the aimed magnetic configurations under the applied current.

For the simulations of equilibrium magnetic configurations, the original Landau-Lifshitz-Gilbert equation is modified by including the iDMI in the free energy $E$. The RK23 (the Bogacki-Shampine version of the Runge-Kutta method) solver is chosen. In each simulation, the solver keeps advancing until the MaxErr, $\epsilon = \max|\tau_{high} - \tau_{low}|\Delta t$ (where $\tau_{high}$ and $\tau_{low}$ are the estimated high-order and low-order torques and $\Delta t$ is the time step), decreases to $10^{-9}$. The initial spin configuration is a numerically conjectured structure, in which a 20 nm wide bubble-like spin texture centered at a site 40 nm far from the nanowire's left edge and 1/4 the wire width far from the top edge is accompanied by a domain wall aligned along the nanowire's central axis.

For the simulations of current-induced dynamics, the conventional LLG equation is extended by the Slonczewski spin-transfer torque. The RK45 (the Dormand-Prince version of the Runge-Kutta method) solver is adopted, and in each simulation, the solver stops advancing when the MaxErr reaches $10^{-5}$. The equilibrium spin configurations obtained from static simulations are used as the input for dynamic simulations.

## APPENDIX B: THEORETICAL MODEL

The Landau-Lifshitz-Gilbert equation is well established as a general-purpose tool for describing the spin dynamics of continuous ferromagnetic systems. From this general equation, the special-purpose Thiele equation can be derived to describe the characteristic of the mediated skyrmion dynamics. Here, to capture the main feature and for simplicity, we regard a skyrmion as a rigid soliton with zero mass, i.e., neglecting the skyrmion's structural deformation during the motion and the skyrmion mass due to the confining potential. In our model, the strip domain wall existing in a nanowire does not manifest itself explicitly in the Thiele equation but enters implicitly into the confining force $\mathbf{F}_p$, and thereby the derivation of the Thiele equation follows that in Refs. [12,30].

**FIGURE CAPTIONS**

**Fig. 1** Domain wall dynamics under various current densities. Panels (a–f) $J$ = 4.0×10$^{12}$ Am$^{-2}$ and panels (g–l) $J$ = 7.0×10$^{12}$ Am$^{-2}$. The time elapsed from current action is indicated in each plot. The arrows in green, magenta, and yellow denote the electric current direction, the spin polarization direction of electrons, and the magnetization distribution in the strip domain wall, respectively. $K_u$ = 0.7 MJm$^{-3}$ and $D$ = 3.0 mJm$^{-2}$. Below $J$ = 4.0×10$^{12}$ Am$^{-2}$, the application of a current has no effect on the strip domain wall, but when $J$ > 4.0×10$^{12}$ Am$^{-2}$, the strip domain wall becomes unstable because of the large-angle precession of the magnetization in the magnetic domains triggered by the spin torque. The color scale is used throughout this paper.

**Fig. 2** Skyrmion motion along a nanowire without including strip domain wall. Two situations are considered: one corresponds to small current density and the other to big current density. Panels (a–f) $J$ = 1.0×10$^{11}$ Am$^{-2}$ and panels (g–l) $J$ = 2.4×10$^{11}$ Am$^{-2}$. The time elapsed from current action is indicated in each plot. The arrows in green denote the electric current. $K_u$ = 0.7 MJm$^{-3}$ and $D$ = 3.0 mJm$^{-2}$. At small current densities, the skyrmion moves through the entire length of the nanowire and stops in the right terminal, whereas at big current densities, the skyrmion moves through only a short distance and then is expelled from the side edge of the nanowire.

**Fig. 3** Skyrmion motion along a nanowire including strip domain wall. Two situations are considered: one corresponds to small current density and the other to big current density. Panels (a–f) $J$ = 1.0×10$^{11}$ Am$^{-2}$ and panels (g–l) $J$ = 4.8×10$^{11}$ Am$^{-2}$. The time elapsed from current action is indicated in each plot. The arrows in green, magenta, and yellow denote the electric current direction, the spin polarization direction of electrons, and the magnetization distribution in the strip domain wall, respectively. $K_u$ = 0.7 MJm$^{-3}$ and $D$ = 3.0 mJm$^{-2}$. At small current densities, the skyrmion steadily slides along the domain

wall string and eventually stops near the right terminal, whereas at big current densities, the skyrmion gradually shrinks during sliding along the domain wall string and finally vanishes when its radius reduces to zero.

**Fig. 4** Skyrmion velocity versus current density. The solid and empty symbols correspond to skyrmion motion in magnetic nanowires with and without including a strip domain wall, respectively. The lines across symbols are only guides to eyes. A series of ($K_u$, $D$) are considered and the plots are shown in panels (a)-(h), respectively. Panel (i) plots the critical skyrmion velocity against critical current density, where the skyrmion is annihilated.

**Fig. 5** Force balance on a steadily flowing skyrmion. (a) Skyrmion motion along the strip domain wall. (b) Skyrmion motion along the sample boundary. $F_g$, $F_D$, $F_{ST}$, and $F_p^\perp$ represent the Magnus, dissipative, driving, and confining forces, respectively. $J$ is the current density and $V_d$ is the skyrmion drift velocity, i.e., $v_x = V_d$. In panel (a), an extra longitudinal repulsive force $F_p^\parallel$ is exerted upon the skyrmion by the domain wall. (c) Magnetization distribution between the mediated skyrmion and sample boundary. "SK border", "DW center", and "Boundary" denote the skyrmion border, domain wall center, and sample boundary, respectively. (d) Magnetization distribution between the bare skyrmion and sample boundary. "SK border" and "Boundary" denote the skyrmion border and sample boundary, respectively.

**Fig. 6** Comparison of the simulation and theoretic results of $\Delta v_x = v_x(\text{mSK}) - v_x(\text{bSK})$ as function of current density. Different combinations of $K_u$ and $D$ are considered as indicated in each panel from (a) to (e). The simulation results of $\Delta v_x$ are derived from the data in Fig. 4. The parameters used for the fittings are summarized in the Supplemental Tables S1–S5 [49].

Fig. 1

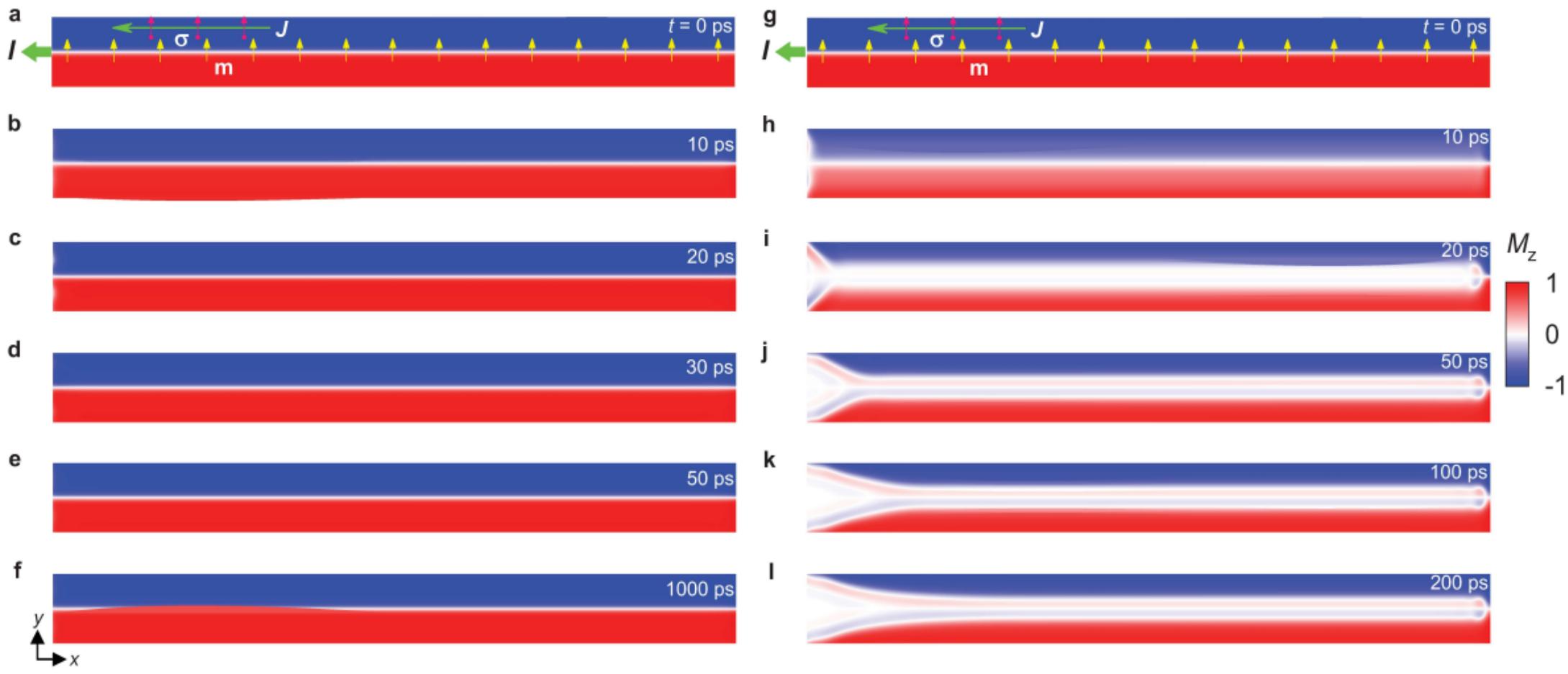

Fig. 2

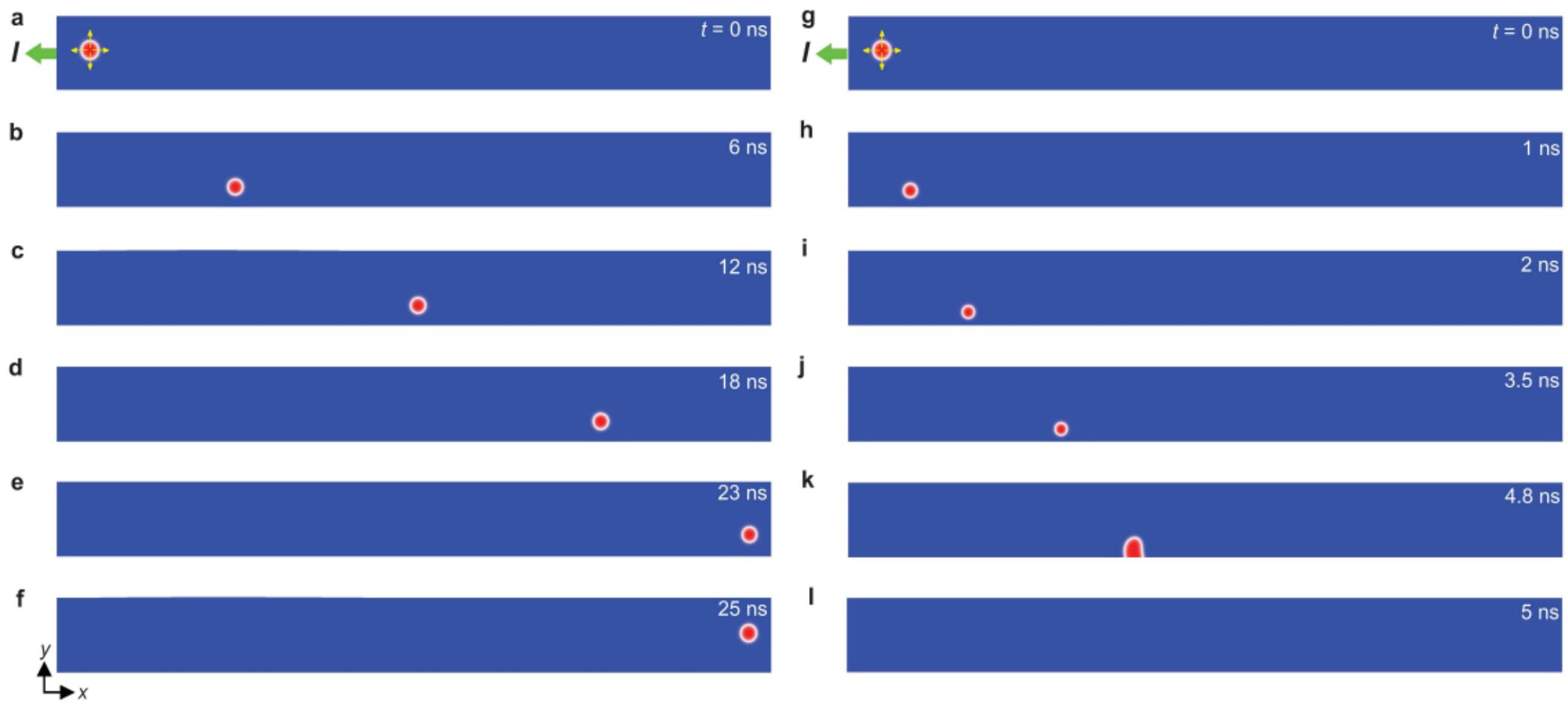

Fig. 3

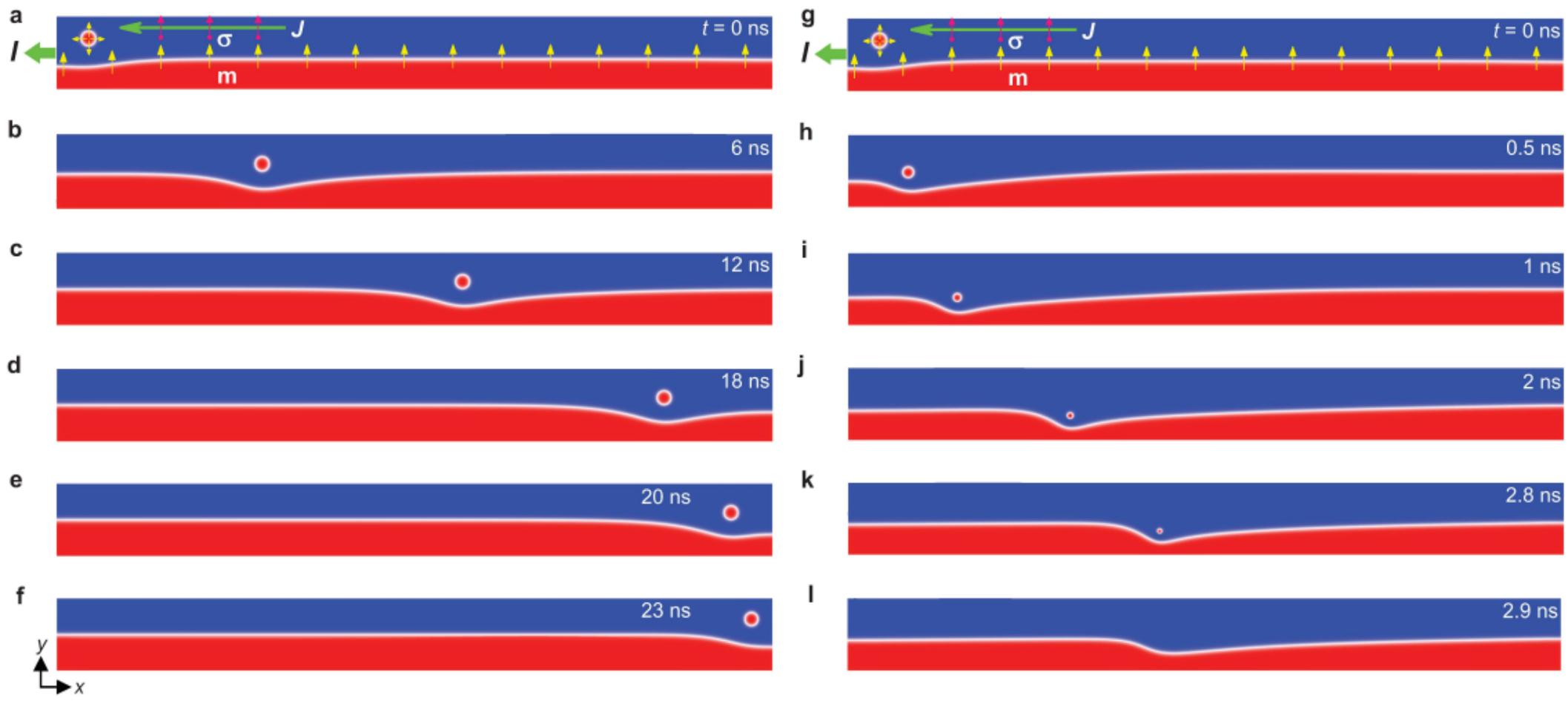

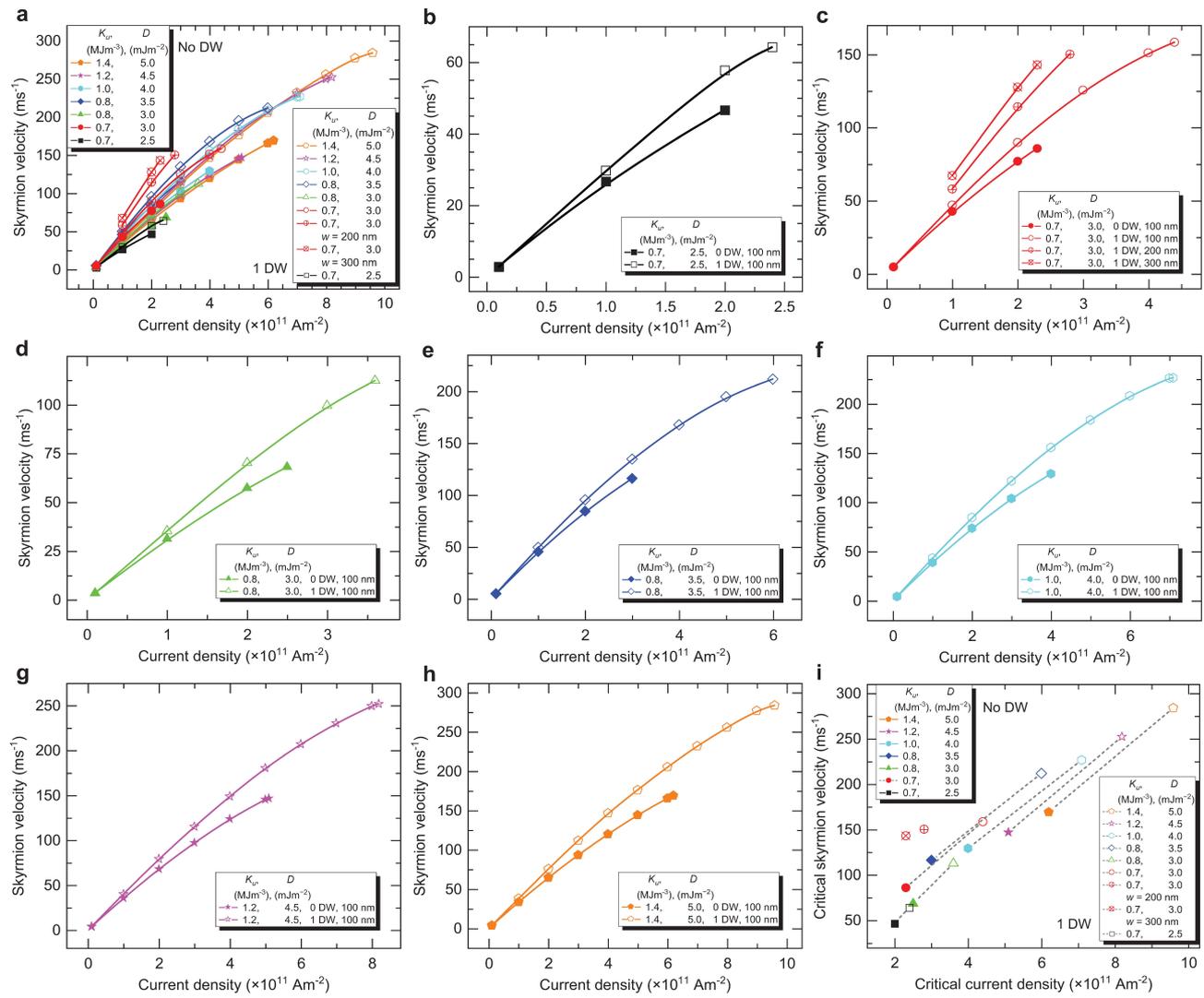

Fig. 5

**a**

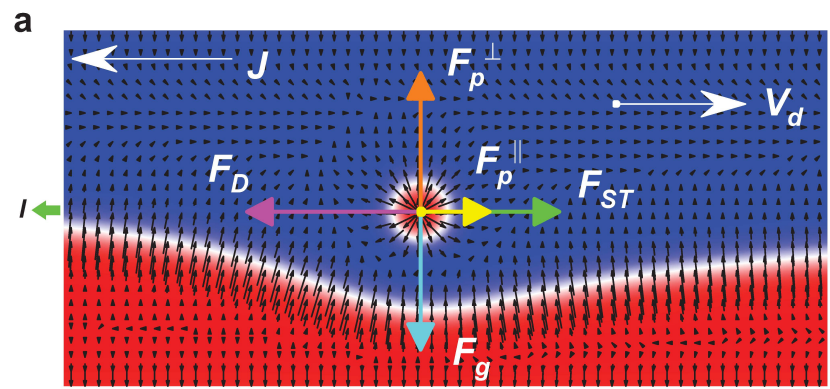

**b**

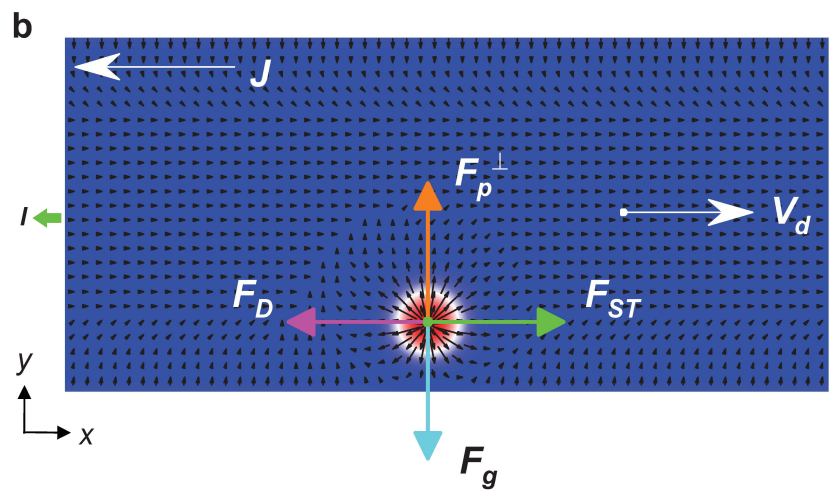

**c**

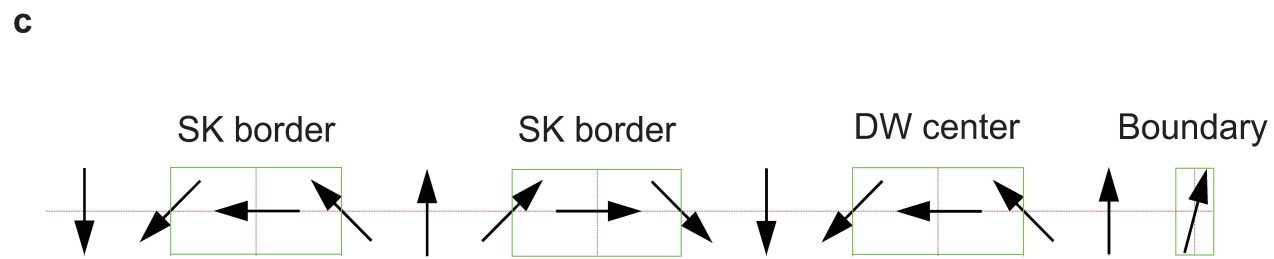

**d**

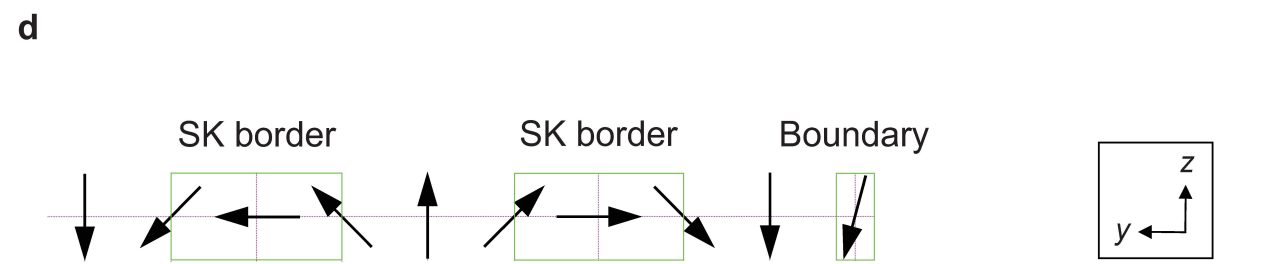

Fig. 6

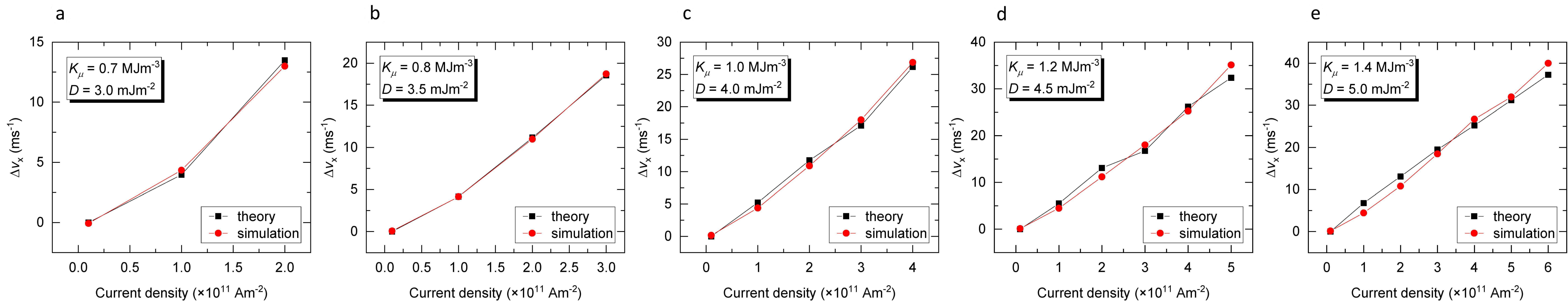

# Supplemental Material for

# Enhanced skyrmion motion via strip domain wall

Xiangjun Xing,[1,*] Johan Åkerman,[2,3,†] and Yan Zhou[4,‡]

[1]School of Physics & Optoelectronic Engineering, Guangdong University of Technology, Guangzhou 510006, China

[2]Department of Physics, University of Gothenburg, Fysikgränd 3, 412 96 Gothenburg, Sweden

[3]Material & Nano Physics, School of ICT, KTH Royal Institute of Technology, 164 40 Kista, Sweden

[4]School of Science & Engineering, The Chinese University of Hong Kong, Shenzhen, Guangdong 518172, China

**This PDF file includes:**

─────────────────────

[*]xjxing@gdut.edu.cn
[†]johan.akerman@physics.gu.se
[‡]zhouyan@cuhk.edu.cn

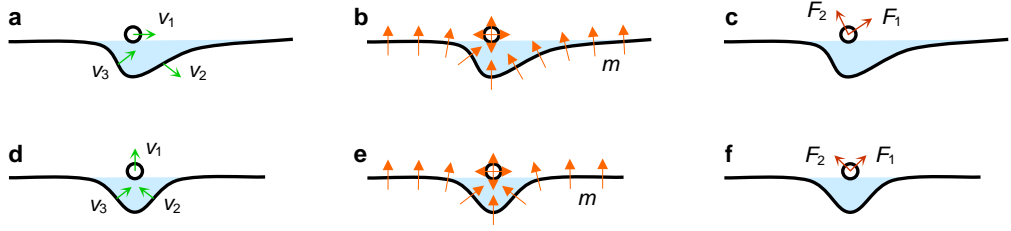

**Fig. S1 Forming mechanism of asymmetric domain wall distortion.** In steadily flowing condition (a), the skyrmion has a constant velocity, $v_1 = V_d$ ($V_d$ is indicated in Fig. 5). At every instant, the front of the bent wall undergoes a thrust force from the moving skyrmion obtaining a velocity $v_2$, whereas the back of the bent wall has a velocity $v_3$ because of the restoring motion driven by the spin orbit torque and the domain wall's own elasticity. As a result, the front and the back of the bent wall become asymmetric. Owing to the mutual repulsion between the magnetic moments on the skyrmion periphery and those on the asymmetrically distorted domain wall (b), the skyrmion experiences two forces: $F_1$ and $F_2$, which are exerted by the back and the front of the bent wall, respectively (c). The net force of $F_1$ and $F_2$ has a transverse component $F_p^{\perp}$, which balances the Magnus force $F_g$, and a longitudinal component $F_p^{\parallel}$, which assists the current-induced force $F_{ST}$ ($F_p^{\perp}$, $F_p^{\parallel}$, $F_g$, and $F_{ST}$ are marked in Fig. 5). If the skyrmion is stationary or moves transversely, the front and the back of the bent wall will keep stationary or move symmetrically (d). Then, the corresponding magnetization distribution will exhibit mirror symmetry (e). Accordingly, in this situation, the net force of $F_1$ and $F_2$ on the skyrmion has only a transverse component (f).

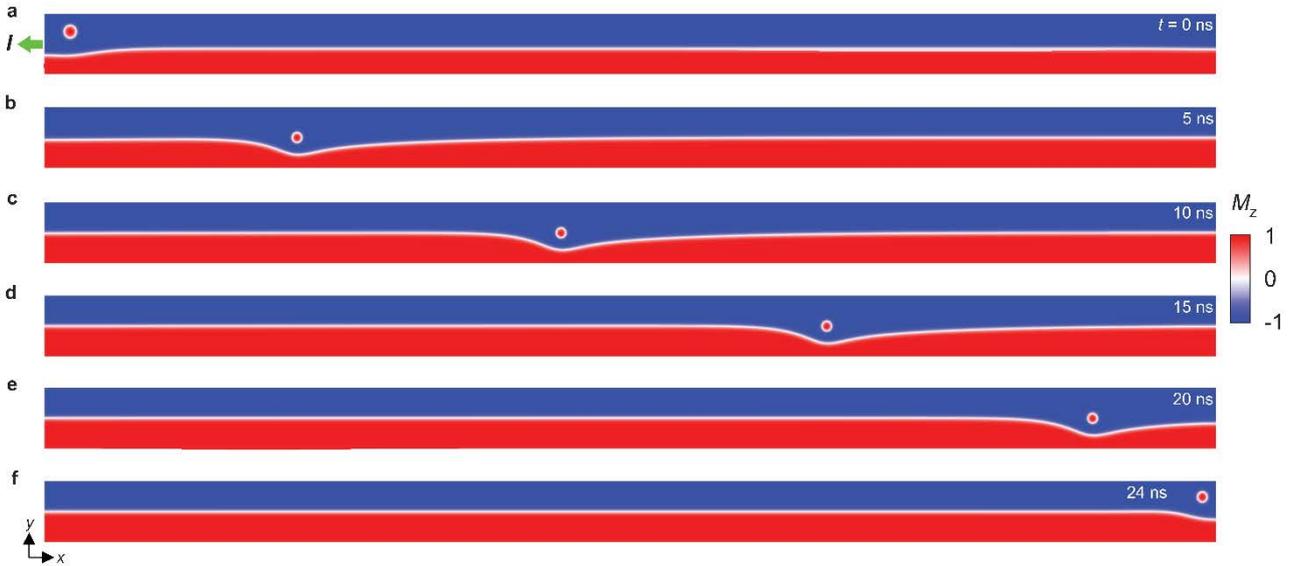

**Fig. S2 Steady skyrmion motion in a longer nanowire including a strip domain wall.** The nanowire is 2 μm in length. The current density $J = 2.0 \times 10^{11}$ Am$^{-2}$. The time elapsed from current action is indicated in each panel of (a–f). The arrow in green denotes the electric current direction. $K_u = 0.7$ MJm$^{-3}$ and $D = 3.0$ mJm$^{-2}$.

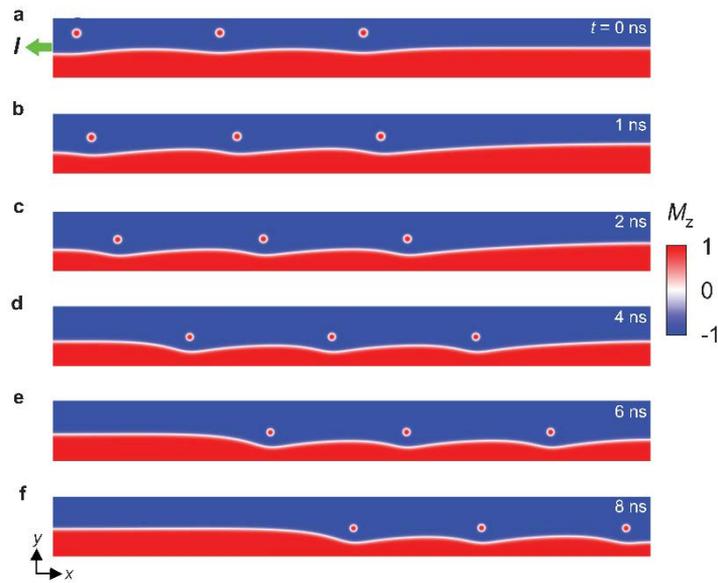

**Fig. S3 Motion of a skyrmion array along a strip domain wall.** The current density $J$ = 2.0×10$^{11}$ Am$^{-2}$. The time elapsed from current action is indicated in each panel of (a–f). The arrow in green denotes the electric current direction. $K_u$ = 0.8 MJm$^{-3}$ and $D$ = 3.0 mJm$^{-2}$. In the beginning, the spacing between two neighboring skyrmions is 240 nm. The skyrmions in an array can move in the same pace only if they are placed enough far from each other or the current density is sufficiently small.

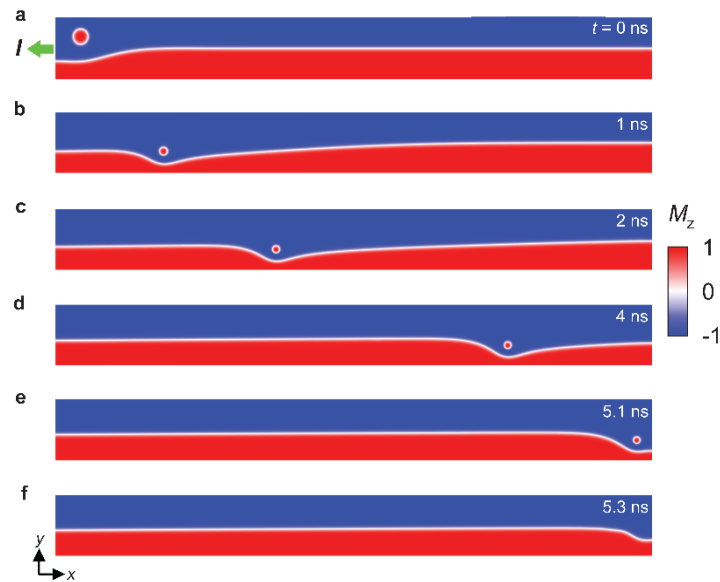

**Fig. S4 Motion of a mediated skyrmion with a considerably large static size.** $K_u$ = 0.8 MJm$^{-3}$ and $D$ = 3.5 mJm$^{-2}$. The current density $J$ = 5.0×10$^{11}$ Am$^{-2}$. The arrow in green denotes the electric current direction. The time elapsed from current action is indicated in each panel of (a–f). The skyrmion slides steadily along the domain wall string and finally is expelled from the right terminal of the nanowire.

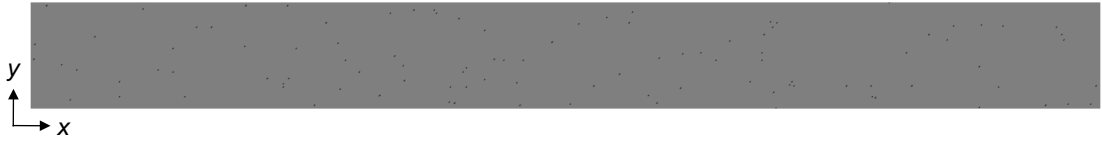

**Fig. S5 Impurity distribution in the magnetic layer.** Randomly distributed point-like defects are considered to mimic the pinning centers ubiquitously existing in practical ultrathin magnetic multilayer samples. The point-impurity concentration is 0.1%. The fluctuating interface-sensitive magnetic parameters, such as the magnetocrystalline anisotropy and the DMI, are the primary source of ordinary pinning centers. Therefore, in our simulations, the cells representing the impurity sites are supposed to have opposite DMI constant. The mediated skyrmion motion for this impurity model is examined against a set of discrete current densities; the supplementary Movies S1–S3 show the processes at three different current densities (In each movie, the time elapsed from current action is indicated. $K_u$ = 0.7 MJm$^{-3}$ and $D$ = 3.0 mJm$^{-2}$).

**TABLE SI** The values of $\eta$, $\eta_0$, and $d_{SK}$ derived from numerical simulations corresponding to the driving current density $J$ for $K_u$ = 0.7 MJm$^{-3}$ and $D$ = 3.0 mJm$^{-2}$. $\lambda_{DW}$ = 5.5 nm is calculated directly from the used material parameters. $\zeta$ = 1.275 ns$^{-1}$.

| $K_u$ = 0.7 MJm$^{-3}$, $D$ = 3.0 mJm$^{-2}$ | | | |
|---|---|---|---|
| $J$ (×10$^{11}$ Am$^{-2}$) | 0.1 | 1 | 2 |
| $\eta$ (nm) | 37.5 | 33.5 | 27.5 |
| $\eta_0$ (nm) | 37.5 | 37.5 | 37.5 |
| $d_{SK}$ (nm) | 19 | 19 | 14 |

**TABLE SII** The values of $\eta$, $\eta_0$, and $d_{SK}$ derived from numerical simulations corresponding to the driving current density $J$ for $K_u$ = 0.8 MJm$^{-3}$ and $D$ = 3.5 mJm$^{-2}$. $\lambda_{DW}$ = 5.0 nm is calculated directly from the used material parameters. $\zeta$ = 1.57 ns$^{-1}$.

| $K_u$ = 0.8 MJm$^{-3}$, $D$ = 3.5 mJm$^{-2}$ | | | | |
|---|---|---|---|---|
| $J$ (×10$^{11}$ Am$^{-2}$) | 0.1 | 1 | 2 | 3 |
| $\eta$ (nm) | 39 | 34.5 | 29 | 25 |
| $\eta_0$ (nm) | 39 | 39 | 39 | 39 |
| $d_{SK}$ (nm) | 24 | 23 | 19 | 16 |

**TABLE SIII** The values of $\eta$, $\eta_0$, and $d_{SK}$ derived from numerical simulations corresponding to the driving current density $J$ for $K_u$ = 1.0 MJm$^{-3}$ and $D$ = 4.0 mJm$^{-2}$. $\lambda_{DW}$ = 4.3 nm is calculated directly from the used material parameters. $\zeta$ = 1.625 ns$^{-1}$.

| $K_u$ = 1.0 MJm$^{-3}$, $D$ = 4.0 mJm$^{-2}$ | | | | | |
|---|---|---|---|---|---|
| $J$ (×10$^{11}$ Am$^{-2}$) | 0.1 | 1 | 2 | 3 | 4 |
| $\eta$ (nm) | 37 | 31.5 | 26.5 | 23.5 | 20.5 |
| $\eta_0$ (nm) | 37 | 37 | 37 | 37 | 37 |
| $d_{SK}$ (nm) | 21 | 20 | 17 | 15 | 12 |

**TABLE SIV** The values of $\eta$, $\eta_0$, and $d_{SK}$ derived from numerical simulations corresponding to the driving current density $J$ for $K_u$ = 1.2 MJm$^{-3}$ and $D$ = 4.5 mJm$^{-2}$. $\lambda_{DW}$ = 3.9 nm is calculated directly from the used material parameters. $\zeta$ = 2.0 ns$^{-1}$.

| $K_u$ = 1.2 MJm$^{-3}$, $D$ = 4.5 mJm$^{-2}$ | | | | | | |
|---|---|---|---|---|---|---|
| $J$ ($\times 10^{11}$ Am$^{-2}$) | 0.1 | 1 | 2 | 3 | 4 | 5 |
| $\eta$ (nm) | 35.5 | 30.5 | 25.5 | 23.5 | 20.5 | 18.5 |
| $\eta_0$ (nm) | 35.5 | 35.5 | 35.5 | 35.5 | 35.5 | 35.5 |
| $d_{SK}$ (nm) | 20 | 19 | 16 | 15 | 12 | 11 |

**TABLE SV** The values of $\eta$, $\eta_0$, and $d_{SK}$ derived from numerical simulations corresponding to the driving current density $J$ for $K_u$ = 1.4 MJm$^{-3}$ and $D$ = 5.0 mJm$^{-2}$. $\lambda_{DW}$ = 3.5 nm is calculated directly from the used material parameters. $\zeta$ = 2.45 ns$^{-1}$.

| $K_u$ = 1.4 MJm$^{-3}$, $D$ = 5.0 mJm$^{-2}$ | | | | | | | |
|---|---|---|---|---|---|---|---|
| $J$ ($\times 10^{11}$ Am$^{-2}$) | 0.1 | 1 | 2 | 3 | 4 | 5 | 6 |
| $\eta$ (nm) | 35 | 29.5 | 25.5 | 22.5 | 21 | 19 | 17.5 |
| $\eta_0$ (nm) | 35 | 35 | 35 | 35 | 35 | 35 | 35 |
| $d_{SK}$ (nm) | 21 | 19 | 17 | 15 | 13 | 12 | 11 |